\documentclass{asaetr}
\usepackage{amsfonts}
\usepackage{graphicx}

\input{tcilatex}
\begin{document}

\title{Labeled\ Random Finite Sets vs.\\
Trajectory Random Finite\ Sets}
\author{Ronald Mahler, Eagan MN, USA, EML: mahlerronald@comcast.net}
\maketitle

\begin{abstract}
The paper \cite{MahSensors1019Exact} discussed two approaches for
multitarget tracking (MTT): the generalized labeled multi-Bernoulli (GLMB)
filter and three Poisson multi-Bernoulli mixture (PMBM) filters. \ The paper 
\cite{Mah20220arXiv-TRFS} discussed two frameworks for multitarget
trajectory representation---labeled random finite set (LRFS) and set of
trajectories (SoT)---and the merging of SoT and PMBM into trajectory PMBM
(TPMBM) theory. \ This paper summarizes and augments the main findings of 
\cite{MahSensors1019Exact}, \cite{Mah20220arXiv-TRFS}---specifically, why
SoT, PMBM, and TPMBM are physically and mathematically erroneous.
\end{abstract}

\section{\protect\small 1. Introduction}

In what follows, \textquotedblleft p.\textquotedblright , \textquotedblleft
c.\textquotedblright , \textquotedblleft l.\textquotedblright ,
\textquotedblleft S.\textquotedblright\ abbreviate \textquotedblleft
pages\textquotedblright , \textquotedblleft column\textquotedblright ,
\textquotedblleft lines\textquotedblright , \textquotedblleft
Section,\textquotedblright\ respectively.

1.1 \textit{LRFS}. \ The RFS IDs/labels model appeared in 1997 in \cite[p.
135, 196-197]{GMN} and subsequently in \cite[S. 14.5.6]{Mah-Artech}; was
systematically expanded into LRFS theory in the 2011 conference paper \cite%
{Vo-ISSNIP12-Conjugate} and 2013 paper \cite{VoVoTSPconjugate}; which has
since been widely adopted. \ The first general exact closed-form (ECF)
approximation\footnote{%
In the sense of \cite{MahSensors1019Exact}.} of any version of the
multitarget Bayes recursive filter (MTBRF)---the GLMB filter---also appeared
in \cite{Vo-ISSNIP12-Conjugate}, a surprising discovery that has been widely
adopted or emulated. \ The first PMBM (PMBM-1) filter followed in the 2012
conference paper \cite{WilliamsFUSION12}.\footnote{%
It was actually a PMB filter and thus not ECF, see S. 8.} \ The first use of
Gibbs sampling in RFS (and probably in MTT) was the 2015 GLMB paper \cite%
{HoangVo-arXiv2015}, which is being increasingly adopted or emulated
(including in PMBM in 2017 \cite{FatimiTSP2017-PMBM-Gibbs}). \ 

The latest Gibbs-based GLMB filter implementations can simultaneously track
over a million 3D targets in real time in significant clutter using
off-the-shelf computing equipment\ \cite{Beard2020}, another surprising
development. \ Also, GLMB-type filters have: \ quantifiable approximation
errors \cite{VoVoTSPconjugate}; linear complexity in the number of
measurements \cite{VoVoTSPconjugate}; log-linear complexity in the number of
hypothesized tracks \cite{ShimVo-TSP2023}; and linear complexity in the
number of scans in the multi-scan case \cite{VoTSPMultiscan2019}.

1.2 \textit{SoT}. \ SoT was, nevertheless, specifically devised to supplant
LRFS. \ It was proposed in 2014 in \cite{Svensson-FUSION2014} and elaborated
in 2020 in \cite{JoseTAES2020T-RFS}, to wit: \textquotedblleft the main
purpose of this paper is to establish the theoretical foundations to perform
MTT using sets of trajectories\textquotedblright\ (p. 1687, c. 1), subject
to the proviso that \textquotedblleft a full Bayesian methodology to MTT
should not rely on pragmatic fixes\textquotedblright\ (p. 1689, c. 1). \
Implementations of an approximate TPMBM filter were reported in 2020 in \cite%
{JoseTSP2020T-RFS}. \ 

In \cite{Svensson-FUSION2014}, \cite{JoseTAES2020T-RFS}, SoT was claimed to
be necessary because LRFS is supposedly fundamentally erroneous: \ target
labels are \textquotedblleft artificial\textquotedblright\ and
\textquotedblleft do not represent an underlying physical
reality\textquotedblright\ \cite[p. 3, c. 1]{Svensson-FUSION2014}. \ It was
further claimed that SoT, unlike LRFS, provides physically correct and
comprehensive modeling of multitarget trajectories. \ However, the 2022
paper \cite{Mah20220arXiv-TRFS} has demonstrated that SoT is based on
fundamental mathematical and physical errors, compounded by multiple ad hoc
fixes---see Sections 4-6, 15-17.

1.3 \textit{PMBM}. \ There are, as demonstrated in 2019 in \cite%
{MahSensors1019Exact} (which has been viewed 2500+ times), actually three
successive versions, all of them theoretically erroneous. \ PMBM-3 was
promoted as state-of-the-art\ in 2018 in \cite[p. 222, c.2]%
{MeyerProcIEEE2018}, as was PMBM-2 in 2018 in \cite[p. 1883, c. 1]%
{JoseTAES2018}. \ Yet in \cite{Svensson-FUSION2014}, \cite{JoseTAES2020T-RFS}%
, and \cite{JoseTSP2020T-RFS}, PMBM-1 and not PMBM-2,3 was, without
explanation, chosen for use in TPMBM. This is presumably because SoT and
PMBM-2,3 are mutually contradictory---see Sections 8-14.

What follows are concise summaries of LRFS, SoT, Poisson RFSs (PRFSs),
PMBM-1,2,3, and TPMBM. \ 

\section{\protect\small 2. Labeled Random Finite Sets}

LRFS was introduced in \cite{Vo-ISSNIP12-Conjugate}, \cite{VoVoTSPconjugate}%
. \ The state of a multitarget population at time $\ t$ \ \ is modeled as a 
\textit{labeled finite set} (LFS) 
\begin{equation}
X=\{(x_{1},l_{1}),...,(x_{n},l_{n})\}\subseteq \mathbb{X}_{0}\times \mathbb{L%
}
\end{equation}%
where \ $x_{1},...,x_{n}\in \mathbb{X}_{0}$ \ are the targets' kinematic
states and the \textit{distinct} labels \ $l_{1},...,l_{n}$ $\in \mathbb{L}$
\ \textit{uniquely identify} them. \ Denote the class of LFSs as \ $\mathbb{F%
}_{0}$. \ A label is a symbol for a discrete state variable: \ target
identity (ID) \cite[p. 2, S. II-B]{Mah20220arXiv-TRFS}. \ If \ $X\notin 
\mathbb{F}_{0}$ \ \ then \ $X$ \ is physically impossible---e.g., \ $%
\{(x_{1},l),(x_{2},l)\}$ \ with \ $x_{1}\neq x_{2}$. \ An LRFS is a random
variable (RV) on \ $\mathbb{F}_{0}$ (and thus labels are unknown random
state variables). \ The GLMB\ filter is an ECF approximation of the MTBRF on
\ $\mathbb{F}_{0}$.\footnote{%
The Bernoulli filter \cite{BaTuong-Thesis}, \cite[S. 14.7]{Mah-Artech}, \cite%
{RisticTutorial2013} and \textquotedblleft dyadic filter\textquotedblright\ 
\cite{Mahler-IETDyad-2022} are non-approximate special cases of the MTBRF
on\ $\ \mathbb{F}_{0}$ \ when, respectively, target number cannot exceed 1
or 2.}

A time-evolving multitarget population is a time sequence \ $%
X_{1:i}:X_{1},..,X_{i}\in \mathbb{F}_{0}$ \ at the measurement-collection
times \ $t_{1},...,t_{i}$. \ The \textquotedblleft $l$-trajectory%
\textquotedblright\ of a target with label \ $l$ \ is the time-sequence \ $%
X_{l}^{k}=X_{k}\cap (\mathbb{X}_{0}\times \{l\})$ \ for \ $1\leq k\leq i$,
where \ $|X_{l}^{k}|\leq 1$ \ since \ $X_{k}\in \mathbb{F}_{0}$ \cite[Eq. 4]%
{Mah20220arXiv-TRFS}.\ \ Note that if \ $X_{l}^{i}\neq \emptyset $ \ then \ $%
X_{l}^{i}=\{(x,l)\}$ \ for some \ $x\in \mathbb{X}_{0}$. \ The
time-consecutive nonempty subsequences of an $l$-trajectory are its
\textquotedblleft track segments.\textquotedblright\ \ \ 

Let \ $X_{l}^{k},...,X_{l}^{k+i-1}$ \ be an \ $l$-segment of length \ $i$ \
with start and stop times \ $t_{k},t_{k+i-1}$. \ If we concatenate \ $%
t_{k+i-1}$ \ to each element of \ $X_{l}^{k+i-1}=\{(x^{i},l)\}$ \ for \ $%
i=1,...,k$ \ then an \ $l$-segment can be re-notated as a vector:%
\begin{equation}
T_{l}=((x^{1},l,t_{k}),(x^{2},l,t_{k+1}),...,(x^{i},l,t_{k+i-1})).
\end{equation}

Let \textquotedblleft $A\approx B$\textquotedblright\ abbreviate the phrase
\textquotedblleft $A$ is notationally equivalent to $B$\textquotedblright\ \
(in the sense that $\ A$ \ and \ $B$ \ are characterized by the same
parameters). \ Then the \ $l$-segment can be successively re-notated as
follows:

\begin{eqnarray}
T_{l} &\approx &(l,t_{k},x^{1},..,x^{i})\approx (l,t_{k},x^{1:i}) \\
&\approx &(l,k,x^{1:i})\approx (l,k_{l},x_{l}^{1:i_{l}})  \label{eq-Renotate}
\end{eqnarray}%
where \ $(l,k_{l},x_{l}^{1:i_{l}})$ \ (redundantly) emphasizes the fact that
\ $k_{l}$, \ $i_{l}$, \ and \ $x_{l}^{1},...,x_{l}^{i_{l}}$ \ define the
particular trajectory segment \ $T_{l}$ \ of the particular target \ $l$.

\section{\protect\small 3. Sets of Trajectories\ }

SoT was introduced in\ \cite{Svensson-FUSION2014} and elaborated in \cite%
{JoseTAES2020T-RFS}. \ The sequence $X_{1:i}:X_{1},..,.X_{i}$ \ is replaced
by a SoT $\ \mathbf{T}=\{T_{1},...,T_{n}\}$ \ with distinct elements $\
T_{1},...,T_{n}$, \ where a \textquotedblleft trajectory\textquotedblright\
\ \ $T$ \ is a vector: 
\begin{equation}
T=(k,x^{1:i})=((t_{k},x^{1}),(t_{k+1},x^{2})...,(t_{k+i-1},x^{i})).
\label{eq-Renotate-2}
\end{equation}%
Here, \ $t_{k}$ \ and \ $t_{k+i-1}$ \ are the trajectory's beginning and end
times and \ $x^{1:i}:x^{1},x^{2},...,x^{i}\in \mathbb{X}_{0}$ are its
kinematic states at times \ $t_{k},t_{k+1},...,t_{k+i-1}$. \ A
\textquotedblleft trajectory RFS\textquotedblright\ (TRFS) is an RV whose
realizations are SoTs.

\section{\protect\small 4. The Mathematical Fundamental SoT Error}

Because \ $T_{1},...,T_{n}$ \ are distinct, \ $T_{j}=(k_{j},x_{j}^{1:i_{j}})$
\ for some $k_{j}$, \ $i_{j}$, and \ $x_{j}^{1},...,x_{j}^{i_{j}}$ \ that
are \textit{uniquely associated} with \ $j=1,...,n$. \ That is: \ $T_{j}$ \
has been implicitly assigned a unique integer LRFS label \ $j\in \{1,...,n\}$%
. \ Notational precision thus requires that \ $%
T_{j}=(j,k_{j},x_{j}^{1:i_{j}})$. \ Comparing this with Eq. (\ref%
{eq-Renotate}) we immediately see that SoT arises from (a) a failure to
realize that the \ $j$ \ are mathematically obligatory, and (b) a resulting
decision to ignore (strip) them. \ It follows that SoT is mathematically
fundamentally erroneous.

Moreover, suppose that the \ $j$ \ are not stripped (resulting in
\textquotedblleft labeled SoT\textquotedblright\ or \textquotedblleft
LSoT\textquotedblright ). \ Then it is clear from Eqs. \ref{eq-Renotate},\ref%
{eq-Renotate-2}) that LRFS trajectory segments and LSoT trajectories
mathematically differ only by a change of notation \cite[p. 4, S. III-A]%
{Mah20220arXiv-TRFS}. \ It follows that LSoT is just LRFS expressed in
different notation. \ \ \ \ 

\section{\protect\small 5. SoT Trajectory Modeling Errors}

The simple counterexamples (CE's) in \cite[S. III-D]{Mah20220arXiv-TRFS}
show that this \textquotedblleft $j$-error\textquotedblright\ results in
seriously erroneous modeling of multitarget systems. \ Specifically, \ SoT
allows impossible SoTs (CE-3) and cannot model two common trajectory types:
\ spawned targets (CE-3) and reappearing targets/tracks (CE-4). \ Thus SoT
is, contrary to claim, neither a physically correct nor a physical
comprehensive trajectory model.

\begin{enumerate}
\item \textit{CE-4}: \ Targets can reappear in a scene. \ Moreover, the
output of MTTs and tracker-classifiers often includes tracks that are
dropped and reacquired. \ Since ground truth trajectories must be compared
to estimated trajectories, any comprehensive trajectory model must encompass
such tracks. LRFS does but SoT cannot. \ Consider \ $\mathbf{T}%
=\{T_{1},T_{2}\}$ \ where \ $T_{1}=(k,x^{1:5})$ \ and \ $%
T_{2}=(k+10,y^{1:5}) $. \ There is serious \textquotedblleft tracking
uncertainty\textquotedblright\ because \ $\mathbf{T}$ \ could be a single
reappearing target but SoT forces it to be two consecutive targets. \
Despite the contrary claim in \cite[p. 3, c. 1]{Svensson-FUSION2014},
tracking uncertainty is eliminated (not increased) by restoring stripped
LRFS labels: \ either \ $T_{1}=(1,k,x^{1:5})$ \ and \ $%
T_{2}=(1,k+10,y^{1:5}) $ (single target) or \ $T_{1}=(1,k,x^{1:5})$ \ and \ $%
T_{2}=(2,k+10,y^{1:5})$ \ (two targets).

\item \textit{CE-1}: \ Point targets have no physical extent and thus can
simultaneously have identical kinematical states. \ Let \ $n$ \ such targets
with labels \ $1,...,n$ \ \ evolve identically during times \ $%
t_{1},...,t_{i}$. \ Then the evolving system is \ $X_{1:i}:X_{1},...,X_{i}$
\ with LFSs \ $X_{k}=\{(x_{k},1),...(x_{k},n)\}$ for \ $k=1,...,i$ \ and \ $%
x_{k}\in \mathbb{X}_{0}$. \ When rewritten in SoT notation with unstripped
labels, \ $X_{1:i}$ \ \ is the same as \ $\mathbf{T}%
=\{(1,1,x^{1:i}),...,(n,1,x^{1:i})\}$. \ If SoT is valid then labels can be
stripped and so \ $\mathbf{T}=\{(1,x^{1:i})\}$: \ a single trajectory rather
than $\ n$ \ of them, a contradiction.

\item \textit{CE-3}: \ Consider \ $\mathbf{T}=\{T_{1},T_{2}\}$ \ where $%
T_{1}=(k,x,x^{1})$ \ and \ $T_{2}=(k,x,x^{2})$ \ with \ $x,x^{1},x^{2}$ \
distinct. \ Then $\ \mathbf{T}$ \ is a physically impossible SoT since a
single target \ $x$ \ at time \ $t_{k}$ \ cannot evolve to two different
states \ $x^{1},x^{2}$ \ at time \ $t_{k+1}$.\footnote{%
Note that it would be specious to argue that erroneous SoTs can be ignored
because they are zero-probability events: \ \textit{every} SoT is a
zero-probability event. \ It would be equally problematic to try to
\textquotedblleft repair\textquotedblright\ SoT by excluding erroneous SoTs
from the definition of a SoT. This would require the (likely impossible)
identification of all possible anomalies, followed by a complete revamping
of SoT densities and integrals and any results based on them. \
Irregardless, this would not alter the fact that SoT as defined in \cite%
{JoseTSP2020T-RFS} is seriously erroneous.} \ Now restore the stripped
labels: \ $T_{1}=(1,k,x,x^{1})$ \ and \ $T_{2}=(2,k,x,x^{2})$. \ Then $\ 
\mathbf{T}$ \ represents a target-spawning event. \ That is, targets 1,2 had
the identical state \ $x$ \ at time \ $t_{k}$, \ at which point they
separated and evolved respectively to \ $x^{1}$ \ and \ $x^{2}$.\ \ 
\end{enumerate}

\section{\protect\small 6. The Physical Fundamental SoT Error}

See \cite[S. II-B]{Mah20220arXiv-TRFS}\ for greater detail. \ The $j$-error
seems to have arisen from the following fundamental physical misconceptions
(all drawn from \cite{JoseTAES2020T-RFS}):

\begin{enumerate}
\item (p. 1678, c. 2) \textquotedblleft with the sequence of sets of labeled
targets, there are infinite representations, as the labeling of the targets
is arbitrary\textquotedblright : \ This is immaterial, because \textit{all} 
\textit{state variables have an infinite number of arbitrary representations!%
}\ \ For example: a position requires the following arbitrary and infinite
representation\ scheme: \ specification of a number base, measurement unit,
coordinate system origin, and coordinate system type.

\item (p. 1687, c. 2) \ \textquotedblleft labels do not represent any
physically meaningful property\textquotedblright : \ This is fundmentally
false. \ The arbitrary symbols (labels) for a position can be assigned in a
unique and \textquotedblleft physically meaningful\textquotedblright\ manner
once these specifications have been chosen. \ Similarly for target IDs and
target labels (which are provisional IDs).

\item (p. 1689, c. 1) \textquotedblleft In practice one can employ pragmatic
fixes\ldots to estimate sensible trajectories\ldots For example, one can use
the dynamic model\textquotedblright : \ This is obviously false.\ \ A
\textquotedblleft dynamic model\textquotedblright\ (Markov density a.k.a.
dynamic prior) is a crucial theoretical feature of Bayesian MTT, not a
\textquotedblleft pragmatic fix\textquotedblright\ \cite[S. 3.5.2]%
{Mah-Artech}. \ Most obviously, it allows a tracker to infer that airplanes
cannot execute instantaneous sharp-angle turns.

\item (p. 1686, c. 1) \textquotedblleft \ldots \lbrack target] labels are
unobservable\textquotedblright : \ This is mistaken. \ Suppose that target
states \ $(p,v,l)$ \ consist of position $\ p$, velocity \ $v$, and \textit{%
unique} label \ $l$, and that the sensor observes only position and is
clutter-free with probability of detection \ $p_{D}=1$. \ Then velocity is
also \textquotedblleft unobservable\textquotedblright\ in this sense. \ Yet
velocity is routinely estimated (inferred). \ To a lesser extent the same is
true of ID. \ An airplane can be inferred to be a jet fighter purely from
kinematics. \ Moreover, labels are usually partially observable. \ When this
sensor observes a set \ $X$ \ of separated targets, its \textquotedblleft
measurement\textquotedblright\ is a set of separated positions. \ The
positions and thus the labels must be distinct and therefore the latter are
not \textquotedblleft arbitrary\textquotedblright\ because \ $X$ \ is an
LFS. \ If otherwise, $\ \ X$ \ would be a physically impossible multitarget
state.
\end{enumerate}

\section{\protect\small 7. Poisson Random Finite Sets}

A PRFS on \ $\mathbb{X}_{0}$ \ has multitarget probability density \ $%
f(X)\propto \prod_{x\in X}D(x)$ \ where $D(x)\geq 0$ \ is a density function
on $\mathbb{X}_{0}$ \ \cite[p. 366]{Mah-Artech}, \cite[p. 98]{Mah-Newbook}.
\ Likewise for a PRFS on \ $\mathbb{X}_{0}\times \mathbb{L}$: \ $f(X)\propto
\prod_{(x,l)\in X}D(x,l)$. \ The latter is physically nonviable since its
realizations can be physically impossible, e.g., \ $\{(x_{1},l),(x_{2},l)\}$
\ with \ $x_{1}\neq x_{2}$. \ Such realizations can be avoided only if \ 
\TEXTsymbol{\vert}$\mathbb{L}|=1$ (the unlabeled case), which results in a
physically erroneous state representation: \ distinct targets have distinct
labels, \textit{independently} of the limitations of the sensors that
observe them.\footnote{%
Thus refuting \cite[p. 1884, c. 1, l. 14-15]{JoseTAES2018}: \
\textquotedblleft ...the usual radar tracking case, in which targets do not
have a unique ID...\textquotedblright\ \ This is a category error: targets' 
\textit{distinctness} (which is innate) is confused with their \textit{%
distinguishability} (which requires an observer/sensor).} \ Thus \textit{%
PRFSs on both \ }$\mathbb{X}_{0}\times \mathbb{L}$\textit{\ \ and \ }$%
\mathbb{X}_{0}$\textit{\ \ do not represent underlying physical reality}.%
\footnote{%
Prior to the 2011 LRFS innovations in \cite{Vo-ISSNIP12-Conjugate},
heuristically labeled RFS filters on \ $\mathbb{X}_{0}$ \ were necessary as
a stopgap to avoid computational intractability, see \cite[p. 3, S. III-A]%
{Mah20220arXiv-TRFS}.} \ 

\section{\protect\small 8. PMBM, Version 1}

This 2012 \textquotedblleft unlabeled\textquotedblright\ (PMBM-1) \ version 
\cite{WilliamsFUSION12} is actually PMB.\footnote{%
The actual PBMB-1 filter apparently appeared in 2015 in \cite%
{WilliamsTAES2015}.} \ Any multitarget population is modeled as a PMB RFS on
\ $\mathbb{X}_{0}$ \ \cite[Eqs. 2,12]{WilliamsFUSION12}. \ This models
\textquotedblleft undetected targets\textquotedblright\ (the
\textquotedblleft P\textquotedblright\ or PRFS part) and \textquotedblleft
detected targets\textquotedblright\ (the \textquotedblleft
MB\textquotedblright\ or MB RFS part). \ At each time-step \ $t_{k}$, \ the
collected measurements in the measurement-set \ $Z_{k}$ \ are assumed to be
from newly-detected targets, and thus each measurement is used to construct
a new Bernoulli (\textquotedblleft B\textquotedblright ) component---i.e., a
new target---of an MBM RFS. \ The MBM RFS is then \cite[p. 1105, c. 1, l.
16-23]{WilliamsFUSION12} approximated as an MB RFS. \ \ \ 

\section{\protect\small 9. PMBM-1 Theoretical Errors}

\begin{enumerate}
\item The PMBMB-1 (but not PMB) filter is an ECF approximation of the MTBRF
on \ $\mathbb{X}_{0}$ \ \cite[S. 4.1]{MahSensors1019Exact}. \ It is not a
theoretically rigorous MTT since it cannot inherently maintain trajectories.

\item The assumption that all measurements arise from newly-detected targets
is a physically erroneous, ad hoc fix.

\item This assumption also implies a non-Bayesian multitarget dynamic prior
\ $f_{k|k-1}(X|X_{k-1},Z_{k})$ \ rather than the usual \ $%
f_{k|k-1}(X|X_{k-1})$ \ or the general \ $f_{k|k-1}(X|X_{k-1},Z_{1:k-1})$ \ 
\cite[Eq. 3.54]{Mah-Artech}.\footnote{%
A closely related issue: the transition of \textquotedblleft undetected
targets\textquotedblright\ to \textquotedblleft detected
targets\textquotedblright\ \cite[p. 13, Item 3]{MahSensors1019Exact}. \ This
should be governed by the general Markov density \ $%
f_{k|k-1}(X|X_{k-1},Z_{1:k-1})$. \ This is impossible because target
detection is governed by the general measurement density \ $%
f_{k}(Z|X_{k},Z_{1:k-1})$ \ \cite[Eq. 3.56]{Mah-Artech}, where \ $X_{k}\sim
f_{k|k-1}(X|X_{k-1},Z_{1:k-1})$ \ and \ $Z_{k}\sim f_{k}(Z|X_{k},Z_{1:k-1})$
\ (and \textquotedblleft $\sim $\textquotedblright\ means \textquotedblleft
random sample drawn from\textquotedblright ).}

\item This prior is also logically impossible: \ how can predicted targets \ 
$X$ \ arise from a measurement-set \ $Z_{k}$ \ \ not yet collected?

\item PMBM distributions grow in size with time and thus must be pruned. \
This is theoretically impossible because\ pruned PMBM distributions are not
valid multitarget (let alone PMBM) distributions. \cite[Eq. 46]%
{MahSensors1019Exact}.
\end{enumerate}

\section{\protect\small 10. PMBM, Version 2}

The \textquotedblleft label-augmented\textquotedblright\ (PMBM-2) filter
appeared in 2015 in \cite{WilliamsTAES2015} to address the fact that the
PMBM-1 filter is not a true MTT. \ It is a PMBM-1 filter defined on \ $%
\mathbb{X}_{0}\times \mathbb{L}$ \ rather than \ $\mathbb{X}_{0}$, where
\textquotedblleft ...track continuity is implicitly maintained in the same
way as in JPDA\ [Joint Probabilistic Data Association] and related methods.
\ This can be made explicit by incorporating a label element into the
underlying state space...\textquotedblright\ \cite[p. 1672, c. 2, l. 7-15]%
{WilliamsTAES2015}. \ Hence the (erroneous) claim that the LRFS paper \cite%
{VoVoTSPconjugate} \textquotedblleft ...shows that the labelled case can be
handled within the unlabeled framework by incorporating a label element in
to the underlying state space\textquotedblright\ \cite[p. 1675, c. 2, l.
16-19]{WilliamsTAES2015}. \ This labeling scheme was carried over into the
2018 sequel paper \cite[p. 1886, c. 1, l. 26-29]{JoseTAES2018}. \ 

\section{\protect\small 11. PMBM-2 Theoretical Errors}

See \cite[S. 4.3]{MahSensors1019Exact} for greater detail.

\begin{enumerate}
\item Targets are created from measurements \cite[Eq. 11]{JoseTAES2018}, so
the dynamic prior is \ $f_{k|k-1}(X|X_{k-1},Z_{k})$.

\item PRFSs are implicitly defined on $\ \mathbb{X}_{0}\times \mathbb{L}$ \
and thus are physically nonviable.

\item How can the labeled framework logically be \textquotedblleft
within\textquotedblright\ (i.e., a special case of) the unlabeled framework?
\ The contrary is true: \ the unlabeled case is $\ |\mathbb{L}|=1$.
\end{enumerate}

\section{\protect\small 12. PMBM, Version 3}

This \textquotedblleft hybrid labeled-unlabeled\textquotedblright\ version
(PMBM-3) was introduced in the 2018 paper \cite{MeyerProcIEEE2018}. \ It is
a modification of the PMBM-2 filter that appears intended to repair the fact
that PRFS's on $\ \mathbb{X}_{0}\times \mathbb{L}$ \ are not LRFSs and thus
cannot model any targets (let alone \textquotedblleft
undetected\textquotedblright\ ones). \ So, detected targets are created from
collected measurements and modeled as labeled MBM (LMBM) distributions, but
now with measurements used as de facto target labels \cite[p. 249, c. 1, l.
8-11]{MeyerProcIEEE2018}. \ Moreover, undetected targets are assumed to have
the same label and thus can be modeled as a PRFS on \ $\mathbb{X}_{0}$.

\section{\protect\small 13. PMBM-3 Theoretical Errors}

See \cite[S. 4.4, p. 13]{MahSensors1019Exact} for greater detail.

\begin{enumerate}
\item It still uses \ $f_{k|k-1}(X|X_{k-1},Z_{k})$.

\item The ad hoc assumption that measurements are de facto target labels
leads to a mathematical contradiction.

\item The assumption that undetected targets have the same label is an ad
hoc fix. \ Its consequence is that such targets can be in multiple locations
simultaneously, a physical impossibility.

\item The PMBM-3 filter does not (as claimed in \cite{MeyerProcIEEE2018})
exactly solve the hybrid labeled-unlabeled MTBRF.
\end{enumerate}

\section{\protect\small 14. SoT Contradicts PMBM-2,3}

\begin{enumerate}
\item Labels are employed in PMBM-2,3 but forbidden in SoT.
\end{enumerate}

\section{\protect\small 15. TPMBM Theoretical Errors}

\begin{enumerate}
\item The TPMBM filter appears to try to repair the errors in the PMBM-2,3
filters by reverting to the PMBM-1 filter and using SoT to enable it to
maintain tracks. \ But this does not alter the fact that PMBM-1 and SoT are
themselves erroneous.

\item Like the PMBM-1 filter, the TPMBM filter presumes that measurements
initiate new targets, to wit: \ \textquotedblleft For a new Bernoulli
component $i$...which is initiated by measurement \ $z_{k}^{j}$%
...\textquotedblright\ \cite[p. 4937, c. 2]{JoseTSP2020T-RFS}.

\item TPMBM requires \textquotedblleft trajectory Poisson
RFSs\textquotedblright\ (TPRFSs): \ that is, PRFSs whose realizations are
SoTs. \ But these are physically nonviable for the same reason that PRFSs on
\ $\mathbb{X}_{0}\times \mathbb{L}$ \ are physically nonviable: they (and
indeed SoT itself) allow physically impossible realizations (see \cite[p. 6,
S. III-E(5)]{Mah20220arXiv-TRFS} and Counterexample CE-3).
\end{enumerate}

\section{\protect\small 16. TPMBM Numerical Errors: TPHD\ Filter}

See \cite[p. 7, S. III-E(4)]{Mah20220arXiv-TRFS} for greater detail. \ The
2019 paper \cite{GarciaSvensson-TSP2019} describes the \textquotedblleft
trajectory probability hypothesis density\textquotedblright\ (TPHD) filter:
\ a first-order approximation of the TPMBM filter. \ Like the TPMBM filter,
it requires erroneous TPRFSs. \ It also employs a direct generalization of
the conventional (unlabeled) PHD filter's multitarget state estimator (which
is summarized in \cite[S. III-E(2)]{Mah20220arXiv-TRFS}). \ The first step
is to find \ $\arg \sup_{T}D(T)$ \ where \ $D(T)\geq 0$ \ \ is a TPHD with \ 
$T=(k,x^{1:i})$. \ If $\ u$ \ is the UoM of \ $\mathbb{X}_{0}$ \ then the
UoMs of \ $T$ \ \ and \ \ $D(T)$ \ \ are \ $u^{i}$ \ and \ $u^{-i}$, which
vary with \ $i$. \ The \ $\arg \sup $ $\ $is therefore mathematically
undefined since the values of \ \ $D(T)$\ \ are numerically incommensurable.%
\footnote{%
This should have been obvious, since one of the earliest RFS\ insights was
that the maximum a posteriori (MAP) estimator is mathematically undefined in
multitarget problems and thus must be replaced by alternatives such as
\textquotedblleft JoM\textquotedblright\ or \textquotedblleft
MaM\textquotedblright\ \cite[p. 59, c. 2]{Mah-Tut}, \cite[S. 14.5]%
{Mah-Artech}, \cite[S. 5.3]{Mah-Newbook}.} \ 

Given this serious numerical error, the favorable simulation results
reported in \cite{GarciaSvensson-TSP2019} require substantive explanation.

\newpage
\includegraphics[width=7in]{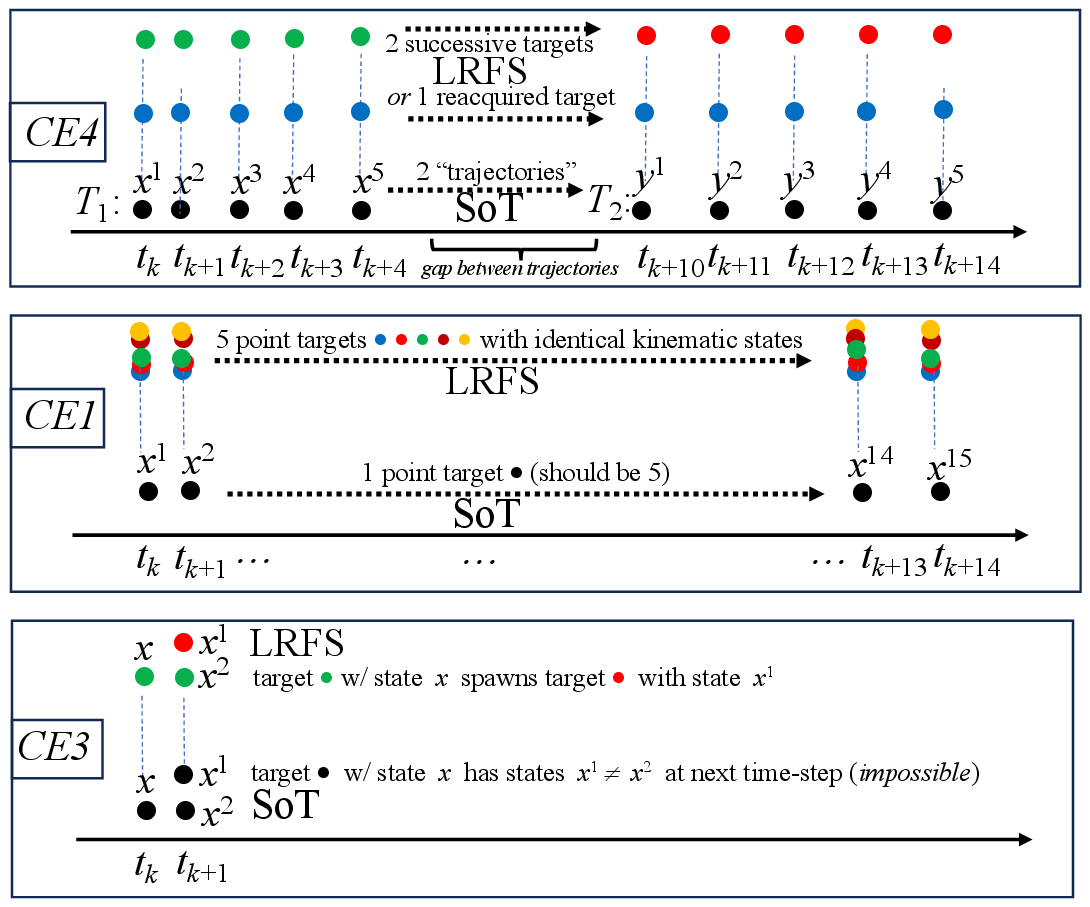}

\end{document}